\documentclass[a4paper]{amsart}
\usepackage[latin1]{inputenc}
\usepackage{amsmath,amsthm,amssymb,latexsym,amsfonts,mathrsfs,enumerate}
\usepackage{tikz-cd}
\usepackage{tkz-tab}
\usepackage{subcaption} 
\usepackage{caption}
\usetikzlibrary{cd}
\usepackage{logicproof}
\RequirePackage{array}
\RequirePackage{ifthen}
\usepackage{accents}
\usepackage{cancel}
\usepackage[foot]{amsaddr}
\usepackage{bibleref}
\usepackage{verse}
\usepackage{bibleref}
\usepackage{color}
\usepackage{dsfont}
\usepackage{amssymb}
\usepackage{graphics}
\usepackage[ND,SEQ]{prftree}
\usepackage{url}
\usepackage[fleqn]{mathtools}
 
\usepackage{amsmath,amsthm,amssymb,latexsym,amsfonts,mathrsfs}
\RequirePackage{mdwtab,latexsym,amsmath,amsfonts,ifthen,pigpen}
\usepackage[usestackEOL]{stackengine}
\usepackage{amssymb}
\usepackage{mathtools}
\usepackage{letltxmacro}
\usepackage{indentfirst}
\usepackage{graphicx}
\usepackage{float}
 \usepackage{mathtools}
 \usepackage{kpfonts}
 \usepackage{enumerate}
 \usepackage{cancel}
  \usepackage{yfonts}
  \usepackage{lettrine}
  \usepackage[ND,SEQ]{prftree}
  \usepackage{imakeidx}
  \graphicspath{ {figures/} }
\usepackage{array}

\theoremstyle{definition}

\theoremstyle{definition}

\numberwithin{Def1}{subsection}
\theoremstyle{definition}

\theoremstyle{theorem}
\newtheorem{theo}{Theorem}
\numberwithin{theo}{section}
\theoremstyle{theorem}

\theoremstyle{theorem}

\theoremstyle{definition}
\newtheorem*{Dem}{\emph{Proof}}
\theoremstyle{definition}

\stackMath
\def\imp{\rightarrow}
\stackMath

\stackMath
\def\ex{\exists}
\stackMath

\stackMath

\stackMath

\stackMath

\def\implic{\twoheadrightarrow}

\makeatletter
\newcommand*\bigcdot{\mathpalette\bigcdot@{.5}}
\newcommand*\bigcdot@[2]{\mathbin{\vcenter{\hbox{\scalebox{#2}{$\m@th#1\bullet$}}}}}
\makeatother

\newlength{\fitchlineht}
\newlength{\fitchindent}
\newlength{\fitchcomind}
\newlength{\fitchnumwd}
\makeatletter
\newcommand\fvline[1][\arrayrulewidth]{\vrule\@height.5\fitchlineht\@width#1\relax}
\makeatother

\newcounter{fitchcounter}
\newboolean{resetfitchcounter}
\setboolean{resetfitchcounter}{true}
\newboolean{increasefitchcounter}
\setboolean{increasefitchcounter}{true}
\newcommand{\formatfitchcounter}[1]{\arabic{#1}}
\newcommand{\fitchcounter}{%
  \ifthenelse{\boolean{increasefitchcounter}}{\addtocounter{fitchcounter}{1}}{}
  \formatfitchcounter{fitchcounter}}

\newenvironment{fitchnum}%
{\ifthenelse{\boolean{resetfitchcounter}}{\setcounter{fitchcounter}{0}}{}
  \begin{tabular}{!{\makebox[\fitchnumwd][r]{\fitchcounter }\hspace{\fitchindent}}Ml@{\hspace{\fitchcomind}}l}}%
{\end{tabular}}

\newenvironment{fitchunum}%
{\begin{tabular}{!{\makebox[\fitchnumwd][r]{}\hspace{\fitchindent}}Ml@{\hspace{\fitchcomind}}l}}%
{\end{tabular}}

\newenvironment{fitch}{
  \begin{fitchnum}}{\end{fitchnum}}
\newenvironment{fitch*}{
  \begin{fitchunum}}{\end{fitchunum}}

{\begin{eqnarray}
    &#1\label{#2}\\
    &\begin{fitch}}%
    {\end{fitch}\notag\end{eqnarray}}

\def\newop#1{\expandafter\def\csname #1\endcsname%
     {\mathop{\rm #1} \nolimits}}
\newop{Con}
\newop{Eq}
\newop{Fix}
\newop{min}
\newop{Ker}
\newop{Ru}
\newop{V}
\newop{Term}
\newop{Sent}
\newop{PA}
\newop{Bew}
\newop{GL}
\newop{Val}
\newop{Var}
\newop{CTerm}
\newop{Form}
\newop{var}
\newop{ZF}
\newop{On}
\newop{Witn}
\newop{K}
\newop{AxT}
\newop{AxY}
\newop{LTL}
\newop{TY}
\newop{Q}
\newop{Diag}
\newop{d}
\newop{f}
\newop{nil}
\newop{term}
\newop{init}
\newop{th}

\renewcommand{\qedsymbol}{\hfill\ensuremath{\blacksquare}}

\tikzcdset{arrow style=tikz, diagrams={>=stealth}}

\begin{document}

\title[]{A Note on a Unifying Proof of the Undecidability of Several Diagrammatic Properties of Term Rewriting Systems}

\author{
  A\MakeLowercase{nt\'{o}nio} M\MakeLowercase{alheiro}\\
  Centro de Matem\'{a}tica e Aplica\c{c}\~oes \& Departamento de Matem\'{a}tica,  F\MakeLowercase{aculdade de }C\MakeLowercase{iências e }T\MakeLowercase{ecnologia, }U\MakeLowercase{niversidade }N\MakeLowercase{ova de }L\MakeLowercase{isboa}\\
  \MakeLowercase{\texttt{ajm@fct.unl.pt}}
  \\
  \\
  P\MakeLowercase{aulo} G\MakeLowercase{uilherme} S\MakeLowercase{antos}\\
  Centro de Matem\'{a}tica e Aplica\c{c}\~oes, F\MakeLowercase{aculdade de }C\MakeLowercase{iências e }T\MakeLowercase{ecnologia, }U\MakeLowercase{niversidade }N\MakeLowercase{ova de }L\MakeLowercase{isboa}\\
  \MakeLowercase{\texttt{pgd.santos@campus.fct.unl.pt}}
}

 \subjclass[2000]{}
\begin{abstract} In this note we give a simple unifying proof of the undecidability of several diagrammatic properties of term rewriting systems that include: local confluence, strong confluence, diamond property, subcommutative property, and the existence of successor. The idea is to code configurations of Turing Machines into terms, and then define a suitable relation on those terms such that the termination of the Turing Machine becomes equivalent to the satisfiability of the diagrammatic property.

\end{abstract}
\thanks{This work was funded by the Funda\c{c}\~{a}o para a Ci\^{e}ncia e Tecnologia through the project PTDC/MHC-FIL/2583/2014 (Hilbert's 24$^{\th}$ problem), through the project PTDC/MAT-PUR/31174/2017 (Semigroups: Conjugacy, Computation, Crystals, and Combinatorics), and trough the project UID/MAT/00297/2019 (Centro de Matem\'{a}tica e Aplica\c{c}\~{o}es).}
\maketitle
\section{Introduction}
Undecidability of important properties of term rewriting systems is a known phenomenon (see \cite{HUET1980349} and, for instance, \cite[p.134]{baader1999term}). See \cite{ENDRULLIS2011227} and \cite{Geser02relativeundecidability} for a logical analysis of the undecidability of confluence. We are going to prove the undecidability of properties related to confluence (strong confluence, diamond property, and many others). We start by presenting the properties about term rewriting systems that we are going to prove to be undecidable, some of them will coincide with known properties\textemdash for instance strong confluence, see figure \ref{fig1}\textemdash and some of them are new properties\textemdash see figure~\ref{fig2}. 

Given a term rewriting systems with rule $\imp$, we say that a property $P(x_0,\ldots,x_n)$ is a \emph{diamond-like property} if 
\begin{align*} P(x_0,\ldots,x_n) \iff \left(\left(\bigwedge_{i=1}^n (x_0\imp x_i) \right)\implies \left(\ex y.~\bigwedge_{i=1}^n (x_i\overset{k_i}{\imp} y)\right)\right),
\end{align*}
where, for each $i\in\{1,\ldots,n\}$, $k_i\in\{*,+,=,\epsilon\}$, where $\epsilon$ denotes the empty word, so $\overset{\epsilon}{\imp}$ is the same as $\imp$. As usual, we say that the term rewriting system has the property $P(x_0,\ldots,x_n)$ if it holds for all terms. There are important examples of diamond-like properties:
\begin{figure}[H]
\centering
\begin{tikzcd}
 \cdot \arrow[r] \arrow[d]
    & \cdot \arrow[d,dashrightarrow,"*"]\\
  \cdot \arrow[r,dashrightarrow,"*"]
&\cdot
\end{tikzcd}
\hspace{1cm}
\begin{tikzcd}
 \cdot \arrow[r] \arrow[d]
    & \cdot \arrow[d,dashrightarrow,"*"]\\
  \cdot \arrow[r,dashrightarrow,"="]
&\cdot
\end{tikzcd}
\hspace{1cm}
\begin{tikzcd}
 \cdot \arrow[r] \arrow[d]
    & \cdot \arrow[d,dashrightarrow,""]\\
  \cdot \arrow[r,dashrightarrow,""]
&\cdot
\end{tikzcd}
\hspace{1cm}
\begin{tikzcd}
 \cdot \arrow[r] \arrow[d]
    & \cdot \arrow[d,dashrightarrow,"="]\\
  \cdot \arrow[r,dashrightarrow,"="]
&\cdot
\end{tikzcd}
\hspace{1cm}
\begin{tikzcd}
 \cdot \arrow[d]
    \\
  \cdot \arrow[r,dashrightarrow,""]
&\cdot
\end{tikzcd}
\caption{From left to right, we have: local confluence \cite[p.28]{baader1999term}, strong confluence \cite[p.28]{baader1999term}, diamond property \cite[p.28]{baader1999term}, subcommutative property, and the existence of successor.}\label{fig1}
\end{figure}
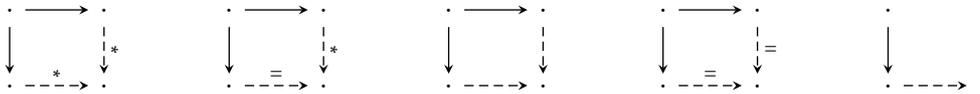
We can even have more complex situations like the following figure illustrates:
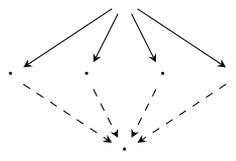
\begin{figure}[H]
\begin{tikzpicture}[scale=.5]
  \node (one) at (0,2) {$\cdot$};
  \node (a) at (-3,0) {$\cdot$};
  \node (b) at (-1,0) {$\cdot$};
  \node (c) at (1,0) {$\cdot$};
  \node (d) at (3,0) {$\cdot$};
  \node (zero) at (0,-2) {$\cdot$};
  \draw [->,>=stealth] (one) -- (a);
  \draw [->,>=stealth] (one) -- (b);
  \draw [->,>=stealth] (one) -- (c);
  \draw [->,>=stealth] (one) -- (d);
  \draw [dashed,->,>=stealth] (a) -- (zero);
  \draw [dashed,->,>=stealth] (b) -- (zero);
  \draw [dashed,->,>=stealth] (c) -- (zero);
  \draw [dashed,->,>=stealth] (d) -- (zero);
\end{tikzpicture}
\caption{An example of another diamond-like property.}\label{fig2}
\end{figure}
It is important to observe that the definition of diamond-like property includes an infinite amount of diagrammatic properties, some of them that were not yet proven to be undecidable. In what follows, we will prove that given a fixed diamond-like property, the problem of knowing whether a given term rewriting system obeys it or not is in general undecidable.
\section{Turing Machines, Coding Them into Terms, and a Relation Between Those Terms}
We recall the definition of a Turing Machine. A \emph{(deterministic) Turing Machine} is a $6$-tuple $\langle Q, \Gamma, \epsilon, q_s, q_e, \delta\rangle$, with a finite set of states $Q$, a finite alphabet $\Gamma$, the blank symbol $\epsilon\in \Gamma$, the initial state $q_s\in Q$, the final state $q_e\in Q$, and the step function $\delta: Q\setminus\{q_e\}\times \Gamma \to Q \times \{\mbox{left},\mbox{right}\}\times\Gamma$. A \emph{configuration} $K$ of a Turing Machine $\mathcal{T}$ is a triple $\langle q,p,b \rangle$, where $q\in Q$ is a state and $p \in \mathbb{Z}$ is a position on the tape $b$. The tape $b$ is a function from $\mathbb{Z}$ to $\Gamma$, where only finitely many memory cells are used, in the sense that $b(p)\neq \epsilon$ only holds for finitely many positions $p$. The initial configuration of the Turing Machine $\mathcal{T}$ is $K^{\mathcal{T}}_S=\langle q_s,0,b_{\text{initial}} \rangle$, where for all $x\in \mathbb{Z}$, $b_{\text{initial}}(x)=\epsilon$. The computation relation $\vdash$ is defined by imposing that $\langle q,p,b \rangle \vdash \langle q',p',b' \rangle$ if, and only if:
\begin{itemize}
\item $q\neq q_e$;
\item $\delta(q,b(p))=\langle q',\mbox{direction},a\rangle$;
\item If $\mbox{direction}=\mbox{right}$, then $p'=p+1$, and $p'=p-1$, otherwise;
\item $b'$ is the tap such that $b'(p)=a$ and, for $x\in \mathbb{Z}\setminus\{p\}$, $b'(x)=b(x)$.
\end{itemize}
We say that a Turing Machine $\mathcal{T}$ \emph{terminates} if there is no infinite sequence $K^\mathcal{T}_S\vdash K_0 \vdash K_1 \vdash \ldots$. The problem of knowing whether a Turing Machine terminates is known to be undecidable (see example 3.2 from \cite[p.43]{homer2011computability}).

Now we move to encode configurations of a given Turing Machine into terms of a term rewriting system. There are several ways to do that, in fact any such way where configurations starting from the initial are encoded into ground terms would work for the proof that we present; nevertheless we will stick to one specific encoding that we consider easier to understand for our purposes. Let us consider a Turing Machine $\mathcal{T}$. For each state $q\in Q$, let us consider a binary-function-symbol $f_q$. Given a configuration $\langle q,p,b\rangle$, the two arguments $\ell$ and $r$ of a term $f_q(\ell,r)$ ought to represent the symbols before (including) and after (excluding) the position $p$ on the tape. In order to represent infinitely many blank symbols appearing to the left and to the right of the tape we use the function symbol $\nil$ of arity $0$. We also use a binary concatenation operator $:$ and interpret $a:b:\nil$ as the sequence of symbols $ab$. The finite alphabet $\Gamma$ is represented by a finite set of constants, so we use $\Gamma\subseteq \Sigma_0$. 

Now we consider the following two rules that allow the insertion of blank symbols:
\begin{align*} f_q(xs,\nil) \implic_\mathcal{T} f_q(xs,\epsilon:\nil); &&  f_q(\nil, ys) \implic_\mathcal{T} f_q(\epsilon:\nil,ys).
\end{align*}

We also add the following rules: If $\delta(q,a)=\langle q',\mbox{left},b \rangle$, then 
\begin{align*}f_q(a:l,x:r) \implic_\mathcal{T} f_{q'}(l,b:x:r)
\end{align*}
and if $\delta(q,a)=\langle q',\mbox{right},b \rangle$, then 
\begin{align*}f_q(a:l,x:r) \implic_\mathcal{T} f_{q'}(x:b:l,r).
\end{align*}

Finally, we consider:
\begin{align*} f_{q_e}(x,y) \implic_\mathcal{T} \term; &&  \init \implic_\mathcal{T} f_{q_s}(\nil,\nil).
\end{align*}

Given a Turing Machine $\mathcal{T}$ and a configuration $K$ of $\mathcal{T}$, let $\mathcal{G}_{\mathcal{T}}(K)$ be the term that codifies the configuration $K$ in the previous term rewriting system. The term rewriting system $\implic_\mathcal{T}$ is very well-known, and we have that\footnote{This is the key-feature for the proof. As mentioned before, as long as configurations starting from the initial one are encoded into ground terms the proof that we present is sound.}
\begin{align*} K \overset{+}{\vdash} K' \iff \mathcal{G}_{\mathcal{T}}(K) \overset{+}{\implic_\mathcal{T}} \mathcal{G}_{\mathcal{T}}(K').
\end{align*}
From the previous equivalence we conclude that $\implic_\mathcal{T}$ is a term rewriting version of $\vdash$. It is importante to observe that if $\init \overset{+}{\implic_\mathcal{T}} t$, then $t$ is a ground term (without variables). Moreover, $\mathcal{T}$ terminates if, and only if, $\init  \overset{+}{\implic_\mathcal{T}} \term$. We are going to create a new term rewriting system from a Turing Machine. Given a Turing Machine, let $\imp_{\mathcal{T}}$ be such that $t \imp_{\mathcal{T}} t'$ if, and only if:
\begin{enumerate}[1.)]
\item $t=\init$, and $\init \overset{+}{\implic_\mathcal{T}} t'$, or
\item $\init  \overset{+}{\implic_\mathcal{T}} \term$, and $\init \overset{+}{\implic_\mathcal{T}} t \overset{*}{\implic_\mathcal{T}} t'$.
\end{enumerate}

\section{Main Theorem}
\begin{theo} Given a fixed diamond-like property $P(x_0,\ldots,x_n)$, the following problem is in general undecidable:
\begin{description}
\item[Instance] A term rewriting-system $\imp$.
\item[Question] Does the system $\imp$ satisfy the property $P$?
\end{description}
\end{theo}

\begin{Dem} Let $P(x_0,\ldots,x_n)$ be a fixed diamond-like property equivalent to
\begin{align*} \left(\left(\bigwedge_{i=1}^n (x_0\imp x_i) \right)\implies \left(\ex y.~\bigwedge_{i=1}^n (x_i\overset{k_i}{\imp} y)\right)\right),
\end{align*}
where, for each $i\in\{1,\ldots,n\}$, $k_i\in\{*,+,=,\epsilon\}$. We are going to prove that the problem of deciding whether a Turing Machine $\mathcal{T}$ terminantes is equivalent to the problem of deciding whether $\imp_{\mathcal{T}}$ has the property $P(x_0,\ldots,x_n)$.

Firstly, let us assume that $\mathcal{T}$ is a terminating Turing Machine and let us prove that $\imp_{\mathcal{T}}$ has the property $P(x_0,\ldots,x_n)$. For that, let us assume $\bigwedge_{i=1}^n (t\imp_{\mathcal{T}} t_i)$. As $\mathcal{T}$ terminates, we have that $\init  \overset{+}{\implic_\mathcal{T}} \term$. As the relation $\imp_{\mathcal{T}}$ is defined only for terms that code configurations that occur in a computation starting from the initial configuration, we have that for all $i \in \{1,\ldots,n\}$, $\init \overset{+}{\implic_\mathcal{T}} t_i $. As $\mathcal{T}$ terminates and for each $i \in \{1,\ldots,n\}$, $\init \overset{+}{\implic_\mathcal{T}} t_i $, necessarily for all $i \in \{1,\ldots,n\}$, $\init \overset{+}{\implic_\mathcal{T}} t_i \overset{*}{\implic_\mathcal{T}} \term$. Thus, $t_i\imp_{\mathcal{T}} \term$. In sum, $\ex y.~\bigwedge_{i=1}^n (t_i\overset{k_i}{\imp_{\mathcal{T}}} y)$, as wanted.

Now, let us assume that $\mathcal{T}$ is a Turing Machine such that $\imp_{\mathcal{T}}$ has the property $P(x_0,\ldots,x_n)$ and let us prove that $\mathcal{T}$ terminates. Let us consider the sequence of configurations starting from the initial configuration: $K^{\mathcal{T}}_S\vdash K_0 \vdash K_1 \vdash \ldots $. Let, for each $i>0$, $t_i=\mathcal{G}_{\mathcal{T}}(K_{i-1})$. We have, for each $i>0$, that $\init \overset{+}{\implic_\mathcal{T}} t_i$. So, $\bigwedge_{i=1}^n (\init \imp_{\mathcal{T}} t_i)$. By hypothesis, we have that $\ex y.~\bigwedge_{i=1}^n (t_i\overset{k_i}{\imp_{\mathcal{T}}} y)$. As by definition the only possibility of occurring $\init$ is in the left side of an application of $\implic_\mathcal{T}$, we conclude from $\bigwedge_{i=1}^n (\init \imp_{\mathcal{T}} t_i)$ that for each $i$, $t_i\neq \init$. Thus, the rules applied in $\ex y.~\bigwedge_{i=1}^n (t_i\overset{k_i}{\imp} y)$ are of the form 2.) of the definition of $\imp_{\mathcal{T}}$, in particular $\init  \overset{+}{\implic_\mathcal{T}} \term$, i.e., $\mathcal{T}$ terminates.

\qedsymbol
\end{Dem}
Although the proof that we presented generalises an infinite amount of diagrammatic properties, it is important to observe that properties like confluence and termination do not fit the scope of the proof: the former has as hypothesis an arbitrary long chain of step\textemdash recall that in the diamond-like properties the hypothesis is just one step\textemdash and the latter is not a diagrammatic property that states the existence of a next step.

Despite the previously mentioned fact, it is clear that, in a term rewriting-system, confluence implies local confluence, hence the undecidability of confluence can be obtained, in an indirect way, from the already proven undecidability of local confluence. More generally, any property that implies local confluence can be proved to be, in a similar way, undecidable. 


\bibliographystyle{alpha}
\bibliography{References}

\end{document}